# Repulsive Casimir Force between Dielectric Planes

By

Karen Windmeier Wetz

Department of Natural Sciences, Manatee Community College, 5840 26th St. W., Bradenton, FL 34207, USA

___

**Abstract**

In 1948 H.B.G. Casimir predicted that an attractive force between two perfectly conducting, neutral plates exists due to changes in the electromagnetic vacuum energy caused by the influence of the plates. In 1956, E.M. Lifshitz derived an extension of Casimir's expression applicable to finite temperatures and arbitrary dielectric constants for two half-spaces with a gap in between them. It is shown in this brief report that, while the Lifshitz formula predicts an attractive force when the gap between the two half-spaces is a vacuum, a repulsive force between the two half-spaces results when certain inequalities between the dielectric constants of the half-spaces and the gap occur. The reason for the repulsive force and possible applications will be considered.

___

Since the time that H.B.G. Casimir derived his formula for the attractive force between two perfectly conducting, neutral plates in 1948 [1]

$$F_C(d) = -\frac{\pi^2}{240}\frac{\hbar c}{d^4}S \qquad (1)$$

(where S is the surface area of each plate and d is the distance of separation between them), there has been increasing interest in extending this derivation to non-perfectly conducting plates, non-zero temperatures and alternative boundary geometries [2]. In his 1956 paper entitled "The Theory of Molecular Attractive Forces" [3], E.M. Lifshitz considered fluctuating electromagnetic fields from zero-point oscillations and thermal sources to derive a formula for the force between two half-spaces with arbitrary dielectric constants and at finite temperatures that reduces to the classical Casimir force in the limit that the dielectric constants of the two half-spaces go to infinity (for perfect conductors) and the temperature goes to zero. The purpose of this brief report is to show that *repulsive* forces can also arise between two dielectric half-spaces under certain conditions.

The geometry of the dielectrics under consideration is given in Figure 1. In 1956, Lifshitz derived an equation for the force between two dielectric half spaces separated by a distance d for the case of a vacuum ($\varepsilon_3 = 1$) in the gap between the two dielectrics [3]. In a 1978 paper by J. Schwinger, L. DeRaad and K. Milton [4] using source theory, an essentially identical formula was derived but without the assumption that $\varepsilon_3 = 1$. The well known result for the force between dielectric half-spaces at zero temperature is [5]:

$$F(d) = \frac{-S\hbar}{2\pi^2 c^3} \int_1^\infty dp\, p^2 \int_0^\infty d\xi\, \xi^3 \varepsilon_3^{3/2} \left( [\frac{\varepsilon_3 s_1 + \varepsilon_1 p}{\varepsilon_3 s_1 - \varepsilon_1 p} \frac{\varepsilon_3 s_2 + \varepsilon_2 p}{\varepsilon_3 s_2 - \varepsilon_2 p} e^{2\xi p \sqrt{\varepsilon_3} d/c} - 1]^{-1} + [\frac{s_1 + p}{s_1 - p} \frac{s_2 + p}{s_2 - p} e^{2\xi p \sqrt{\varepsilon_3} d/c} - 1]^{-1} \right)$$

(2)

where $s_{1,2}^2 \equiv p^2 - 1 + \frac{\varepsilon_{1,2}}{\varepsilon_3}$, S is the area of dielectric plane and $\varepsilon_j = \varepsilon_j(i\xi)$ are the real-valued dielectric constants (j=1,2,3) evaluated at imaginary frequency $i\xi$. Although detailed knowledge of the frequency dependence of each dielectric constant is required before this expression can be evaluated exactly, it is possible to determine some of the properties of F(d) without explicit calculation. For example, it is not hard to show that when there is a vacuum in the central region of space and the other two dielectric constants are greater than one that the force given by equation (2) is attractive. This is because $s_{1,2}$ is greater than p in that case, and so each factor that precedes the exponential function is greater than one, leading to a positive-definite integrand. Most people will not consider the attractive nature of this force surprising because of the well known connection between this force and the attractive retarded Van der Waals (or, Casimir-Polder) forces [5]. It seems to have been overlooked in the literature, however, that the force given by equation (2) can also be repulsive under suitable circumstances, such as when the dielectric constants obey the inequality $\varepsilon_1 > \varepsilon_3 > \varepsilon_2$ in the range of frequencies that contribute most substantially to the integrals. (Other cases of repulsive "Casimir forces" have already been explored for the case of two conducting plates in a Fabry-Perot cavity [6] and for assorted other types of geometries and fields [2]). An expression for the repulsive force will be derived in what follows, along with a physical interpretation and possible applications.

For the sake of an explicit calculation, let us consider the case where the dielectric constant of the first half space is greater than that of the gap by a small amount so that $\varepsilon_1 = \varepsilon_3 + \delta$, where $\delta$ is a small positive quantity, and the dielectric constant of the second half space is slightly less than that of the gap, $\varepsilon_2 = \varepsilon_3 - \delta$, for frequencies of importance (i.e. for those less than or equal to $\frac{c}{d\sqrt{\varepsilon_3}}$). We will define $\Delta = \frac{\delta}{\varepsilon_3}$, where we have assumed $\Delta < 1$, and expand equation (2) in powers of $\Delta$. First, let us rewrite (2) in a more convenient form:

$$F(d) = -\frac{\hbar S}{2\pi^2 c^3} \int_1^\infty dp\, p^2 \int_0^\infty d\xi\, \xi^3 \varepsilon^{3/2} \left( \frac{G}{1-G} + \frac{H}{1-H} \right) \tag{3}$$

where $G = \frac{\varepsilon_3 s_1 - \varepsilon_1 p}{\varepsilon_3 s_1 + \varepsilon_1 p} \frac{\varepsilon_3 s_2 - \varepsilon_2 p}{\varepsilon_3 s_2 + \varepsilon_2 p} \exp(-2\xi p \sqrt{\varepsilon_3}\, d/c)$ and

$H = \frac{s_1 - p}{s_1 + p} \frac{s_2 - p}{s_2 + p} \exp(-2\xi p \sqrt{\varepsilon_3}\, d/c)$.

After expansion and algebraic manipulation of the coefficients of the exponential functions, it is easy to show that, to second order in $\Delta$,

$$G = -\left(\frac{\Delta}{2}\right)^2 \left(1 - \frac{1}{2p^2}\right)^2 e^{-2\xi p \sqrt{\varepsilon}\frac{d}{c}} + \text{h.o.t.}, \text{ and } H = -\left(\frac{\Delta}{2}\right)^2 \left(\frac{1}{2p^2}\right)^2 e^{-2\xi p \sqrt{\varepsilon}\frac{d}{c}} + \text{h.o.t.}$$

(I have dropped the subscript "3" on the dielectric constant $\varepsilon_3$, for simplicity). One can observe at this point that *F(d)* will be a *repulsive* force, since G and H are definitely negative functions. In order to evaluate this expression explicitly, it is convenient to make the assumption that the dielectric constants are approximately independent of frequency near the frequencies that contribute most substantially to this expression. It is also easy to see that the values of G and H are always less than one so that we can expand

$$\frac{G}{1-G} = \sum_{n=0}^{\infty} G^{n+1} \quad \text{and} \quad \frac{H}{1-H} = \sum_{n=0}^{\infty} H^{n+1}.$$ Only the n=0 term in the summation

contributes to second order in $\Delta$ and, after performing the integrations, one finds:

$$F(d) = \frac{\hbar c S \Delta^2 23}{d^4 \sqrt{\varepsilon} \pi^2 2^7 5} + h.o.t. \tag{4}$$

It is not hard to evaluate the correction terms to order $\Delta^4$ as well, although those terms will not be included here. While at first glance it might seem surprising that a repulsive force can arise between two dielectric half-spaces separated by a third dielectric, it makes some intuitive sense if one recalls that Casimir forces are related to the attractive, retarded van der Waals (Casimir-Polder) forces. The net force on dielectric 3 in the gap is then expected to be directed towards the substance with the largest dielectric constant (medium 1 in this case) due to the van der Waals forces and so the substance in the gap would seem to pull towards medium 1 and away from medium 2, resulting in what might appear to be a "repulsive force" (i.e. the distance d of separation of the gap would tend to increase). If we compare the magnitude of this force per unit area to that between two perfectly conducting plates as given by equation (1), we find the ratio:

$$\frac{F(d)}{F_C(d)} \approx \frac{-\Delta^2(.0885)}{\sqrt{\varepsilon}}.$$

Since the Casimir force between perfectly conducting plates is small and difficult to measure (e.g. .013 dynes for plates of area one square centimeter separated by a distance of one micron) [5], it follows that this considerably smaller force would be even harder to determine experimentally. For example, if a thick plate of light flint glass is separated from a thick slab of fluorite with the liquid toluene occupying a gap of width one micron, one could estimate (using $\Delta \approx .09$ and $\varepsilon_3 \approx 2.25$ and ignoring dispersion effects) that the repulsive force has a magnitude of $6.5 \times 10^{-6}$ dynes per square centimeter, which is one twentieth of one percent of the Casimir force between two perfectly conducting plates. Of course, this force has a strong dependence on the width of the gap and can be increased significantly by reducing the size of the gap. Possible applications to the field of nanotechnology and surface wetting come to mind, as well as neurobiology (where medium 3 could relate to the material in the synaptic cleft (for which d is typically less than one micron) and mediums 1 and 2 would relate to the material in the presynaptic cell (axon) and the postsynaptic cell (dendrite), respectively, for instance).

## ACKNOWLEDGMENT

I would like to thank Manatee Community College for providing me with a sabbatical leave during which time this research was undertaken.

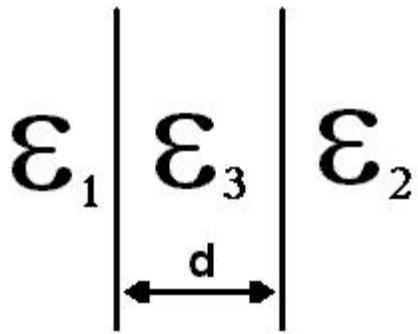

Figure 1. A pair of half-spaces separated by a gap of length d, each filled with media of dielectric constants $\varepsilon_1$, $\varepsilon_2$ and $\varepsilon_3$, respectively.